\documentclass[final,3p,times]{elsarticle}

\usepackage{graphicx}
\usepackage{color}

\usepackage{amssymb}

\newcommand{\g}{\dot{\gamma}}

\journal{Compte rendues}

\begin{document}

\begin{frontmatter}



\title{Viscoelastic surface instabilities}


\author{A. Lindner $^{a}$ and C. Wagner $^{b}$}

\address{$^{a}$ Laboratoire de Physique et M\'ecanique des Milieux
H\'et\'erog\`enes (PMMH), UMR 7636 CNRS - ESPCI - Universit\'es
Paris 6 et 7,  10, rue Vauquelin, 75231 Paris Cedex 05, France\\
$^{b}$ Technische Physik, Universit\"at des Saarlandes, Postfach
151150, 66041 Saarbr\"ucken, Germany}

\begin{abstract}
We review three different types of viscoelastic surface
instabilities: The Rayleigh -- Plateau, the Saffman -- Taylor  and
the Faraday instability. These instabilities are classical examples
of hydrodynamic surface instabilities. The addition of a small
amount of polymers to pure water can alter its flow behavior
drastically and the type of instability may change not only
quantitatively but also qualitatively. We will show that some of the
observed new phenomena can be explained by the use of simple
rheological models that contain most of the underlying physical
mechanisms leading to the instability. A quantitative description however
is often only possible close to the onset of the instability or for
weak deviations from Newtonian behavior. A complete theoretical
description is still lacking when the system is driven far from
equilibrium or for fluids with strong non-Newtonian behavior.

\textbf{R\'esum\'e}

Nous discutons dans ce papier trois types d'instabilit\'es
interfaciales dans des liquides visco-\'elastiques : l'instabilit\'e
de Rayleigh -- Plateau, l'instabilit\'e de Saffman -- Taylor et
l'instabilit\'e de Faraday. Ce sont toutes les trois des exemples
typiques d'instabilit\'es hydrodynamiques de surface. L'addition
d'une faible quantit\'e de polym\`eres \`a de l'eau pure suffit \`a
lui conf\'erer un comportement fortement non-newtonien.  Les
instabilit\'es dans ces solutions sont modifi\'ees pas seulement de
fa\c con quantitative mais aussi dans leur nature. Nous montrons que
des mod\`eles rh\'eologiques simples peuvent expliquer l'origine des
modifications observ\'ees sur les instabilit\'es. Une description
quantitative n'est en g\'en\'eral possible uniquement pr\`es du
seuil d'instabilit\'e et uniquement pour des d\'eviations faibles
d'un comportement newtonien. Une description compl\`ete loin de
l'\'equilibre et pour des propri\'et\'es non-newtoniennes fortes
manque \`a ce jour.

\end{abstract}





\end{frontmatter}



\section{Introduction}
\label{Introduction}

In this paper, we review three typical hydrodynamic surface (or more
general interfacial) instabilities: the Rayleigh -- Plateau
instability, the Saffman -- Taylor instability and the Faraday
instability. These instabilities were chosen as they are each
representative of a more general class of pattern-forming systems
and we have investigated each. Here we discuss the
modifications of these instabilities that occur when using viscoelastic fluids
instead of simple Newtonian liquids.

One of the most prominent examples of a surface instability is
a water droplet that detaches from a faucet (for a review
see \cite{Eggers1997,Eggers2008}). The primary stages of the
detachment of the droplet can be described by the linear Rayleigh --
Plateau instability: a cylindrical thread is destabilized by surface
tension and splits up into individual droplets, thus minimizing
the energy of the system. In the non-linear regime, the final detachment
or pinch-off of the droplet represents a so-called finite
time singularity and must be described in a very different way,
e.g., by use of so-called self similarity solutions. The generation
of droplets also occurs in many industrial applications where,
instead of simple liquids like water or oils, more complex liquids
such as polymer melts, polymer solutions, surfactants, suspensions,
etc. are processed. The hydrodynamics of these liquids differ
from those of Newtonian liquids; as a result, the character of their surface
instabilities can be altered dramatically.
When a tiny amount of a flexible polymer is added to a solvent,
pinch-off is strongly delayed and, in the nonlinear regime, a
viscoelastic filament is formed between the nozzle and the droplet
\cite{Goldin1969, Bazilevskii1981,Renardy1994,Entov1997,Stelter2000,Amarouchene2001,
Anna2001,Wagner2005,Tirtaatmadja2006,Clasen2006}. Interestingly, this contrasts with
theoretical predictions obtained from a linear analysis, in which an increase in
growth rates and an acceleration of the break-off was predicted \cite{Goldin1969}.
Experiments using yield stress fluids \cite{coussot2005,Niedzwiedz2009} or granular
suspensions \cite{Furbank2007} also reveal a strong alteration of the detachment
process.

A classical pattern-forming system containing a free surface is the
Faraday experiment \cite{Faraday1831}. When a layer of liquid is
shaken vertically by a sinusoidal force, the flat interface becomes
unstable as soon as a critical amplitude is exceeded. Standing waves
of high symmetry are formed and patterns of lines, squares,
hexagons, quasiperiodic patterns, superlattices and oscillons have
been reported
\cite{Zhang1996,Lioubashevski1996,Binks1997,Kudrolli1998,Wagner2000}.
Energy is continuously fed into the system by the driving force and
is dissipated by the viscous flow. This leads to a nonequilibrium
situation in which the pattern selection process is no longer described by
thermodynamic laws but only by the nonlinear dynamics of the
system. For higher driving strengths, the pattern dynamics may
become chaotic or weakly turbulent until droplet ejection occurs.
Again, the introduction of a complex liquid can fundamentally alter the pattern
selection process of the system
\cite{Umbanhowar1996,Raynal1999,Wagner1999,Ballesta2005,Kityk2006,Muller1999,Merkt2004}.

Another classical example involving, in this case, a moving interface
is the so-called Saffmann--Taylor instability \cite{Saffman1958}.
When a viscous fluid pushes a less viscous fluid in a narrow channel
or Hele -- Shaw cell, the interface between the two fluids becomes
unstable and the formation of air fingers occurs. This
instability has received much attention as an archetype of pattern-forming
systems; it belongs to the class of free-boundary problems,
in which the growth of the structures takes place in a Laplacian field
\cite{Couder2000}. A linear stability analysis yields a length scale
that in combination with the geometry of the system also determines
the non-linear growth of the viscous fingers observed. The Saffmann
-- Taylor instability has been intensively studied for Newtonian
fluids \cite{Couder1991, Homsy1987,Bensimon1986} and lately also for
non-Newtonian fluids \cite{McCloud1995, VanDamme1989} such as polymer or
surfactant solutions \cite{Lindner2002, Zhao1993, Greffier1998,
Bonn1995} foams, pastes, clays and gels \cite{Lindner2000b,
Puff2002, Park1994, VanDamme1989} and granular suspensions or dry
granular media \cite{Chevalier2006, Chevalier2009, Johnsen2008,
Cheng2008}. The Saffmann-- Taylor instability is important for a number
of applications, including oil
recovery, coatings, and adhesive debonding \cite{Nase2008}. These
applications often involve non-Newtonian fluids, e.g., polymer solutions,
which have been shown to fundamentally alter the selection process of
the viscous fingers (see figure \ref{fig:Pictures}).

Further examples of hydrodynamic surface instabilities include the Kelvin --
Helmholtz instability (the interfacial instability of two liquids that are
sheared against each other), the Rayleigh--Taylor instability (that of
the interface between two liquids of different densities with the
heavier liquid placed on top of the lighter) , the B\'enard Marangoni (an
instability that occurs when a
layer of liquid that is heated from below becomes unstable due to
differences in the surface stresses caused by thermal gradients),
the case in which a thin film flowing down an inclined plane becomes
unstable against wavy distortion \cite{Gupta1967} and the
disintegration of a liquid jet into droplets \cite{Eggers2008}. Some
of these systems have been investigated for the case of complex
liquids. A special case is the so-called shark-skin instability
\cite{Tordella1956,Denn1990}, which refers to the appearance of a wavy
distortion on a polymeric fiber that is forced through a hole or
slit (a "die")  above a critical speed. This instability is attributed
either to an instability at the solid-liquid interface
(stick-slip) or to a bulk instability \cite{Bertola2003}.

For analytical and numerical investigations, treatment of the free interface of these systems is still a
challenge. Despite the fact that these instabilities have been known for a long time, theoretical
understanding of the underlying mechanisms was only obtained
much later. The Faraday Experiment was the first hydrodynamic pattern-forming
system that has been described scientifically\cite{Faraday1831};
however, a successful numerical analysis based on the
full Navier -- Stokes equation was performed only recently
\cite{Perinet2008}. The pinch-off of a water droplet represents a
very general problem; however, it was solved analytically only in the
1990´is \cite{Eggers1993}. The Saffman -- Taylor instability was
described by Saffman and Taylor in 1958 \cite{Saffman1958}, but
the mechanism of finger selection in this instability has remained a puzzle for several decades. The
problem was solved numerically in the '80s \cite{McLean1981}
but was only much later solved analytically \cite{Hong1986,
Shraiman1986, Combescot1986}. Analytical treatment of the case of
viscoelastic liquids remains challenging even today.

In this paper, we discuss experimental modification of the
Rayleigh -- Plateau (followed by the pinch-off of the droplet), the
Saffman -- Taylor and the Faraday instabilities when using one type of
non-Newtonian fluids: polymer solutions. The addition of small
amounts of either flexible or rigid polymers to pure water or water-
glycerol mixtures can dramatically change the properties of the liquid.
We show that the non-Newtonian properties of these
solutions affect the three surface instabilities in very different
ways, and further show that some of the observed new phenomena can be
explained by the use of simple rheological models describing the
hydrodynamics of the complex liquids. However, a quantitative description
is often only possible close to the onset of the instability
or for weak deviations from Newtonian behavior. A complete
theoretical description is still lacking when the system is driven
far from equilibrium or for strong non-Newtonian behavior.

 \section{The hydrodynamics of complex liquids }
\label{HD}

In this chapter, we discuss the hydrodynamics of polymer solutions
and characterize the specific solutions that were used in the
following studies.
From a theoretical point of view, polymers represent
very complex systems due to the large number of degrees of freedom that exist in their structures.
Even a single polymer chain can possess millions degrees of freedom due
to the many independent molecular bonds between monomers that
can rotate more or less freely. In equilibrium, entropic forces
keep the polymer in a mainly coiled state that allows the
highest number of conformations. A first approach to representing polymer
structure is to model the polymer as a series of connected beads. If hydrodynamic interaction
between the beads is taken into account, one obtains the so-called
Zimm model \cite{Zimm1956}. This model predicts that a polymer
that is deformed relaxes to its coiled equilibrium configuration
with a certain relaxation time $\tau_p$. The relaxation
time depends strongly on the molecular weight, the flexibility of
the polymer and the viscosity of the solvent. It is worth pointing
out that the polymer is an entropic spring only, and that in typical
experiments, energies sufficient to affect the bonds are not
reached. In dilute solution, polymers do not
interact with each other. Higher concentrations can be referred
to as semidilute as long as effects such as
reptation modes, which are observed in polymer melts, are not
present.  The hydrodynamics of polymer solutions is challenging
because it is not clear a priori how the coarse graining from the
microscopic polymer models to a continuum mechanical description can
be properly performed. Often, macroscopic models
built on a heuristic base that can give some insight
into the specific flow situation are used.
A quantity that is often used to measure the importance of elastic
effects under flow is the so-called Weissenberg number $Wi=\tau_p \dot{\gamma}$, with $\tau_p$
the relaxation time of the solution and $\dot{\gamma}$ the shear
rate of the flow. It compares in that way the relaxation time of
the polymer to the typical time scale of the flow. Only if the Weissenberg
number is of $\mathcal O(1)$ the flow is strong enough to affect the
polymers significantly.

In our experiments, we deal on the one hand with dilute solutions of
the polymers xanthane and PEO  (polyethylenoxide) and, on the other
hand, with more concentrated solutions of PAA
(polyacrylic-co-acrylacid). These polymers are important for a wide
variety of applications. Xanthane is used to control the rheological
properties of products such as tooth paste, yoghurt, curd and cream; PEO
is used to reduce drag in turbulent flows and PAA is used as a shear
thickener in paper coating applications. These solutions show some
of the most frequently encountered non-Newtonian properties, making
them excellent model systems. We now represent the rheology
of these solutions and discuss their non-Newtonian behavior in detail.

\begin{figure}
  \includegraphics[height=50mm]{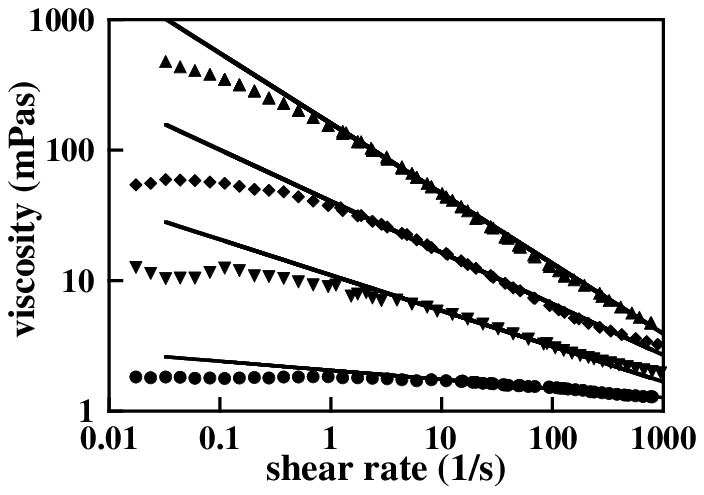}
  \includegraphics[height=50mm]{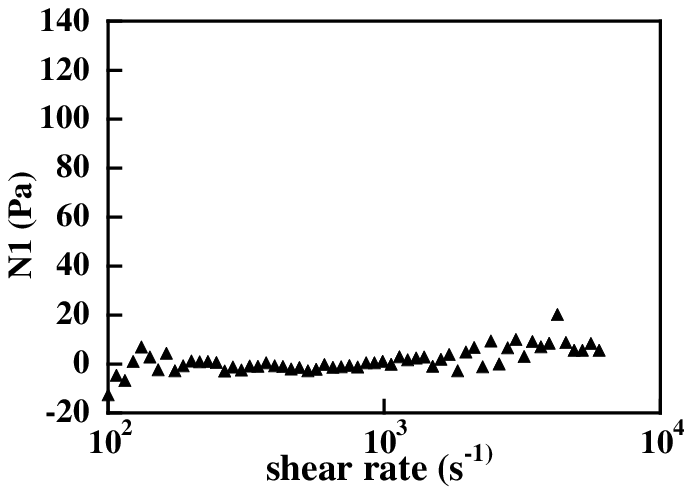}
  \caption{Left) Viscosity $\eta$ as a function of the shear rate
  $\g$ for different concentrations of xanthane, 50~ppm ($\bullet$),
100~ppm ($\blacktriangledown$), 500~ppm ($\blacklozenge$) and
1000~ppm ($\blacktriangle$). Right) Normal stress $N_1$ as a
function of the shear rate $\g$ for a xanthane solution of 1000ppm
($\blacktriangle$).}\label{fig:rheolX}
\end{figure}

Dilute or semi-dilute solutions of the rigid polymer xanthane
($M_w=3*10^6$~g/mol) are shear-thinning. Figure \ref{fig:rheolX}
shows the shear viscosity obtained for solutions of
various concentrations of xanthane. With increasing concentration, an increased
viscosity at low shear rates and increasingly strong shear thinning
when $\g$ is larger than a critical shear rate are observed. On the microscopic
level, one can imagine that the rigid polymers align with the flow at a critical shear rate and
concentration.
The decrease in viscosity can be described by phenomenological models as, e. g., the power law model $\eta=k_1\g^{(n-1)}$, which is valid over a given range of shear rates. The value of $n$ is a measure of the strength of the shear-thinning character of the fluid; for the xanthane solutions, this value ranges from $n$=0.9 for 50 ppm (parts per million in weight) to n=0.46 for 1000 ppm. For $n$=1, the Newtonian limit is recovered. No normal stresses perpendicular to the shear plane can be detected in the tested range of shear rates.

\begin{figure}
\includegraphics[height=50mm]{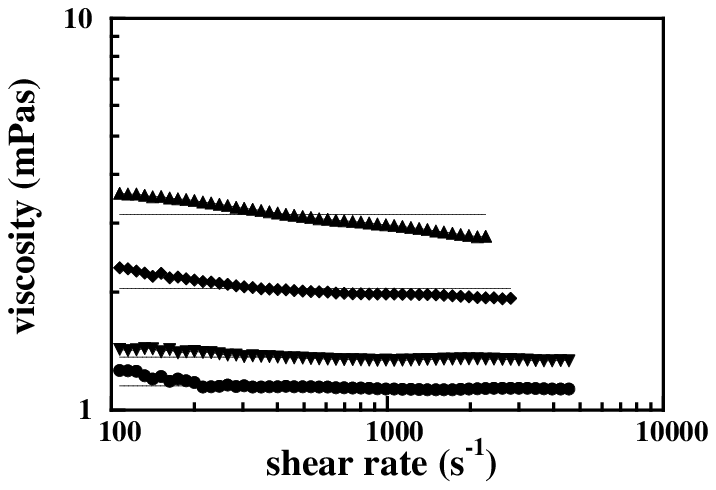}
\includegraphics[height=50mm]{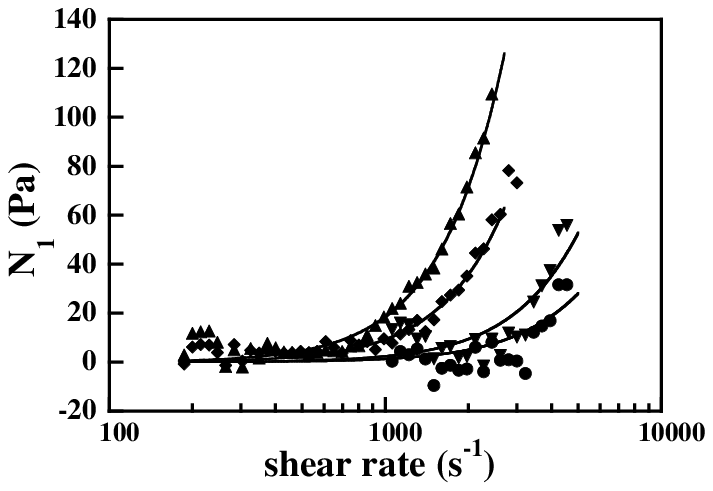}
\caption{(from \cite{Lindner2000a}) Viscosity $\eta$ (left) and
normal stress $N_1$ (right) as a function of the shear rate $\g$ for
solutions of different concentrations of PEO 125~ppm ($\bullet$),
250~ppm ($\blacktriangledown$), 500~ppm ($\blacklozenge$) and
1000~ppm ($\blacktriangle$). The solid lines represent a fit of our
data to the prediction of the Oldroyd-B model.} \label{rheoPEO}
\end{figure}

\begin{figure}
\includegraphics[width=75mm]{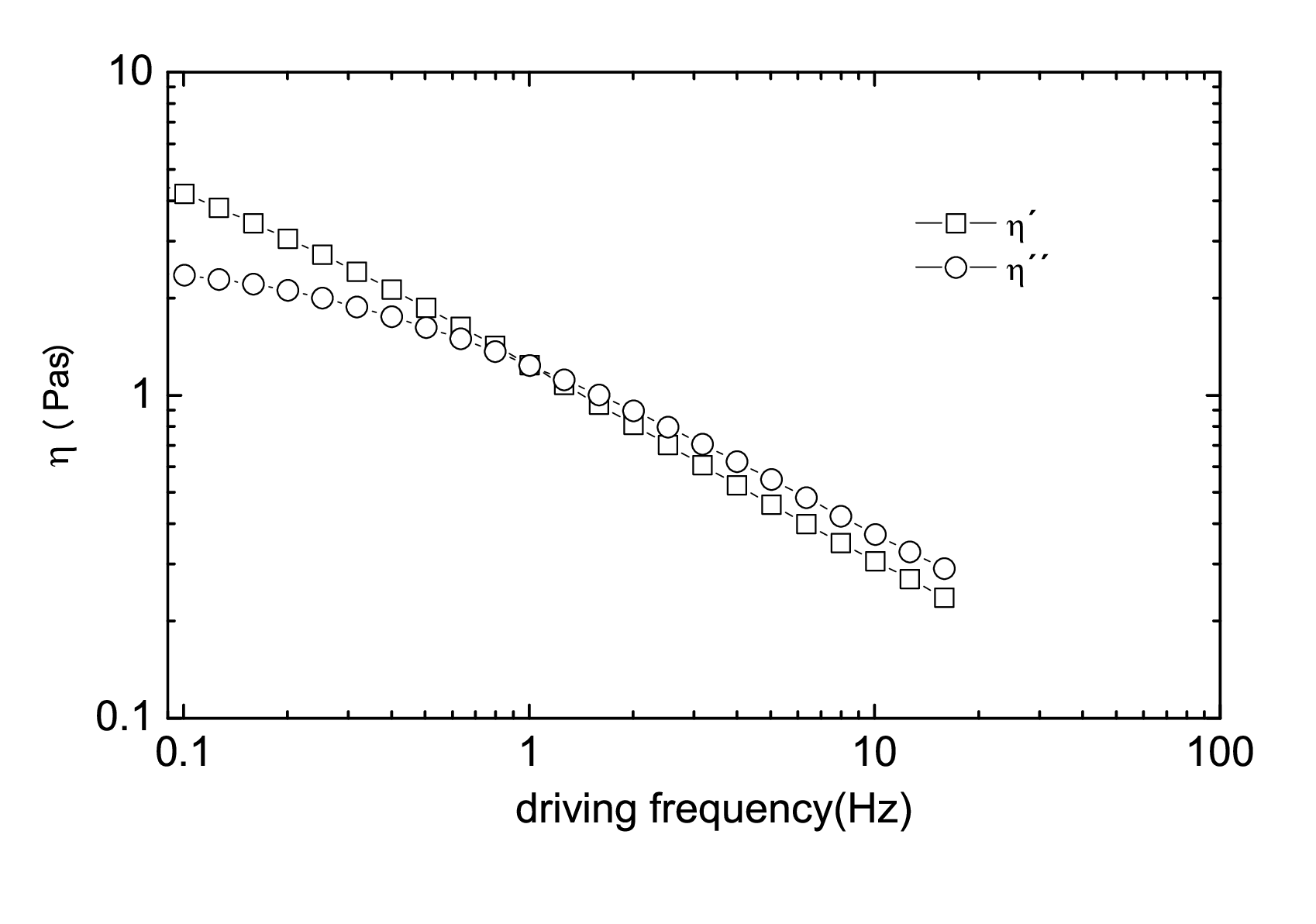}
\caption{Complex viscosity $\eta^*=\eta'+i\eta''$ for a 2000 ppm PAA
($M_w=4*10^6$~g/mol) in $40/60 wt\%$ water glycerol solution. The
intersection of the dissipative $\eta '$ and the elastic part $\eta
''$
 indicate the inverse relaxation time $1/\tau_p$ of the liquid.}
\label{PAAvisco}
\end{figure}

  For the flexible polymer PEO, the shear viscosity is nearly constant, whereas significant normal stresses can be detected (see figure \ref{rheoPEO}).
 With increasing polymer concentration, the Newtonian viscosity increases slightly and the normal stresses become more pronounced.  The normal stress is a stress contribution normal to the shear plane in the direction of the velocity gradient. It is, for example, responsible for the Weissenberg effect: if a rod is used to stir a Newtonian liquid in a cup, centrifugal forces depress the surface in a parabolic form. If large normal stresses are present in a viscoelastic fluid, the inverse can be observed and the liquid climbs the rod, as can be observed by stirring the dough.  The existence of normal stresses also indicates a strong resistance to elongation; thus, elongational viscosities of fluids that undergo normal stress can be orders of magnitude higher than those of Newtonian fluids. For the dilute and semi-dilute polymer solutions presented here, this viscosity is very difficult to obtain by classical rheology \cite{Lindner2003}.

On a microscopic level, one can imagine coils of the flexible polymer that are stretched by the
shearing flow as soon as the shear rate overcomes a critical shear rate, given by the Weissenberg number $Wi$.
The flow properties of solutions of flexible polymers can be modeled by a microscopic bead and dumbbell model in which a polymer is represented by a spring linking two beads. The dumbbells are convected by the flow. Elastic stretching of the dumbbell due to viscous drag and Brownian forces leads to additional stresses in the flow. By averaging and integrating, a continuum mechanical model can be obtained. The simplest of these models is the Oldroyd-B model, in which Hookean springs with infinite extensibility are assumed.  A first outcome of the Oldroyd-B model is indeed the existence of a normal stress.
One finds a constant viscosity and a quadratic increase of the normal stress with shear rate:
$\eta=const$ and $N_1=\Psi_1\g^2$. In this model, the viscosity and $\Psi_1$, the first normal stress coefficient, are independent from the shear rate and are functions of the polymer relaxation time $\tau_p$. On figure \ref{rheoPEO}, one sees that this model fits the experimental data reasonably well. An improved version, the so- called FENE-P model, considers a finite extensibility of the springs and allows description of larger deviations from Newtonian behavior. When used in this manner, the FENE-P model is
also valid for higher shear rates or higher elongation rates.

For concentrated solutions of PAA, one can measure a linear elastic response for low deformation. The macroscopic Maxwell model assumes that the stresses in the liquid are given by
the sum of a viscous part that is proportional to the shear rate and an elastic part that is proportional to the shear or deformation.  The first consequence is that the
viscosity becomes a complex number, $\eta^*=\eta'+i \eta''$, which has both a dissipative and an elastic part and one can again attribute a characteristic relaxation time, $\tau_p$ ,to the
system. The complex viscosity of the solutions as a function of frequency is shown in the figure
\ref{PAAvisco}. One can see that even if the complex viscosity cannot be directly fitted with a
simple Maxwell model, a relaxation time of $\tau_p \simeq 1 s$ can
be extracted from the frequency dependency of the complex viscosity
at the intersection of the $\eta'$ and $\eta''$ curves. Note that
the Oldroyd-B model is also referred to as a convected Maxwell model, since for a
nonlinear, i.e., finite, deformation, the convection  of stress must be taken into account.

\section{Rayleigh -- Plateau instability and droplet pinch-off}
\label{RP}

\subsection{Newtonian liquids}
\label{RP-N}
\begin{figure}
\begin{center}
\includegraphics[width=65mm]{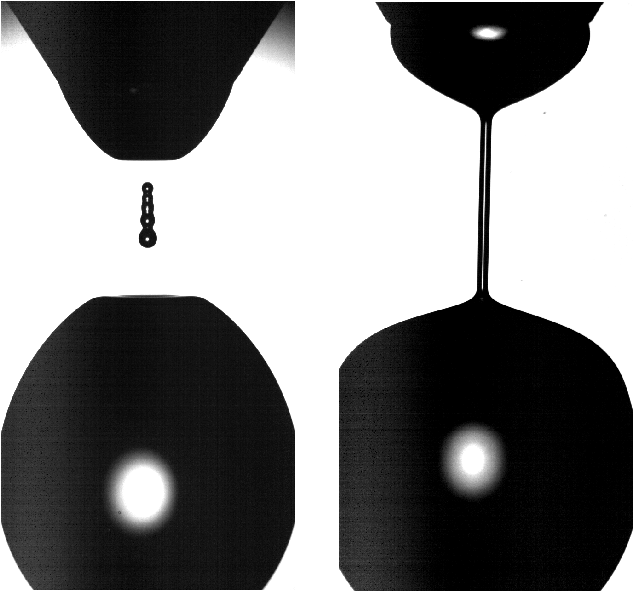}
\end{center}
\caption{Left image: A droplet of water falling from a nozzle. Right
image: The same experiment after the addition of 100ppm PEO
($M_w=4*10^6$~g/mol). } \label{figdroplets}
\end{figure}

A liquid column with a free surface always disintegrates into
smaller droplets because surface tension leads to minimization of the
surface-to-volume ratio. Rayleigh showed that the size $R$ of the
droplets is determined by the wave length $\lambda$ of the
sinusoidal distortion with the fastest growth rate. For the inertia-dominated 
case, i.e., for low viscosity liquids, one finds $\lambda \approx
9 r$ with the radius r of the column. For the pending droplet, a
similar scenario holds. When a pending droplet is fed quasitatically
via a syringe pump, it starts to fall if gravitation overcomes
capillary forces. However, as soon as the droplet begins to fall,
surface tension again leads to minimization of its surface area
and acts as the main pinching force
\cite{Clanet1999,Rothert2003,Wagner2005}. Thus, the primary stages of the
pinch-off of a droplet can be modeled as a Rayleigh instability. The
thinning dynamics of the neck radius, shown in figure
\ref{figdroplets}, can be fitted with an exponential law
corresponding to exponential growth of the amplitude of the most
unstable wavelength. The growth rates fit well with the predictions
made by Rayleigh's theory \cite{Wagner2005}.

During the final stages of the pinch-off process in the nonlinear regime,  the situation becomes
very different. The system no longer reflects the original
geometry, e.g., the diameter of the nozzle, but instead depends on material
parameters (density, surface tension and viscosity) only. The
dynamics of this finite time singularity in which the minimum neck
diameter reaches zero in finite time can be described by self-similarity solutions.
For the low viscosity and thus inertia-dominated regime, which applies in the case of water, it follows that the minimum
neck radius approaches zero such that $r(t) = 0.7 (t-t_c)^{2/3}$ (figure
\ref{figdroplets}) \cite{Eggers1993}.

\subsection{Polymer solutions}

\label{RP-PS}

The effect of elasticity on the (linear) dynamics of the
Rayleigh-Plateau instability can be best studied theoretically by
use of the (linear) Maxwell model.  Chang {\it et al.}
\cite{Chang1999} performed a linear stability analysis and predicted
slight variations of the critical wave numbers and of the growth
rates due to linear elasticity.

\begin{figure}
\includegraphics[width=80 mm]{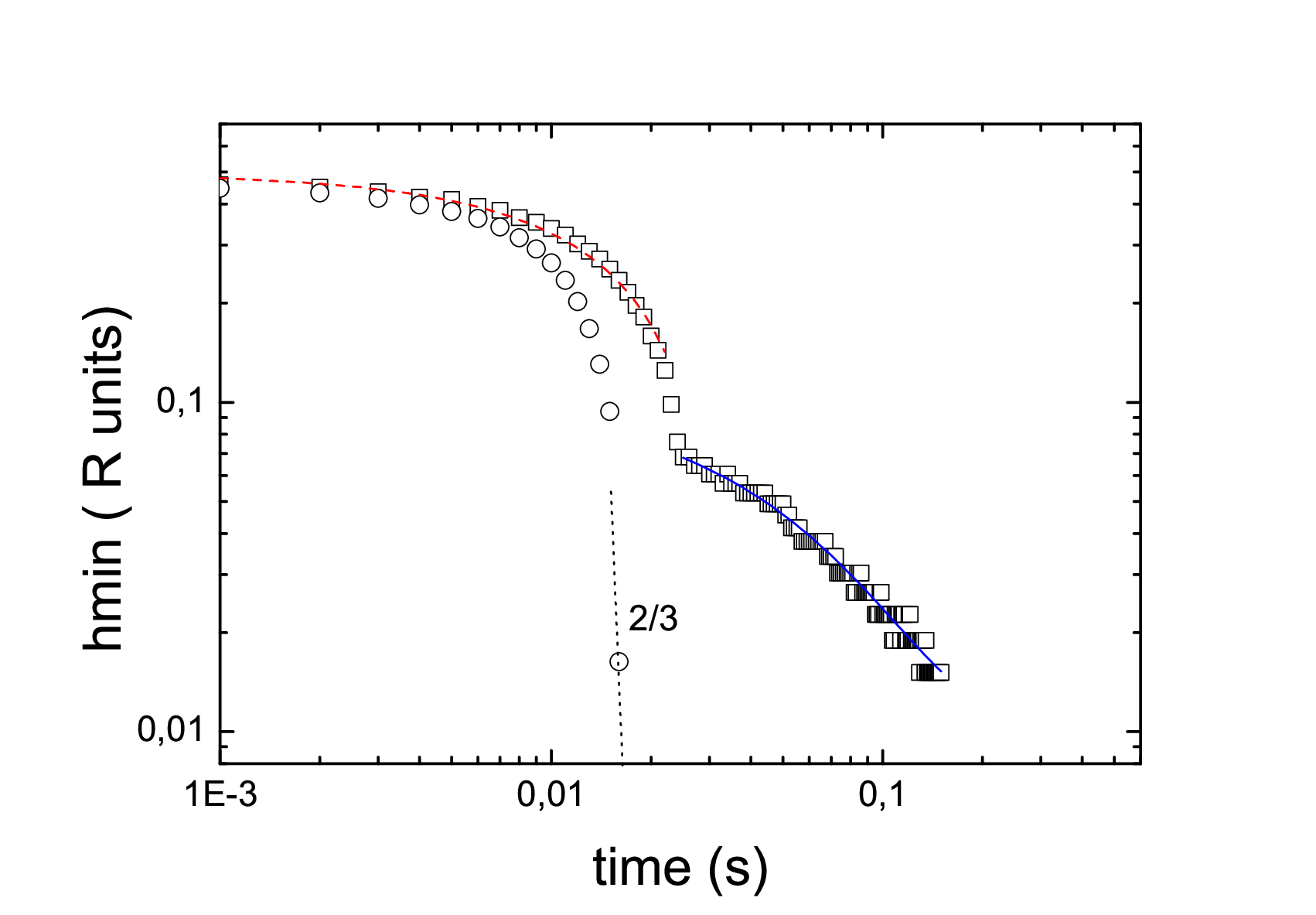}
\includegraphics[width=80mm]{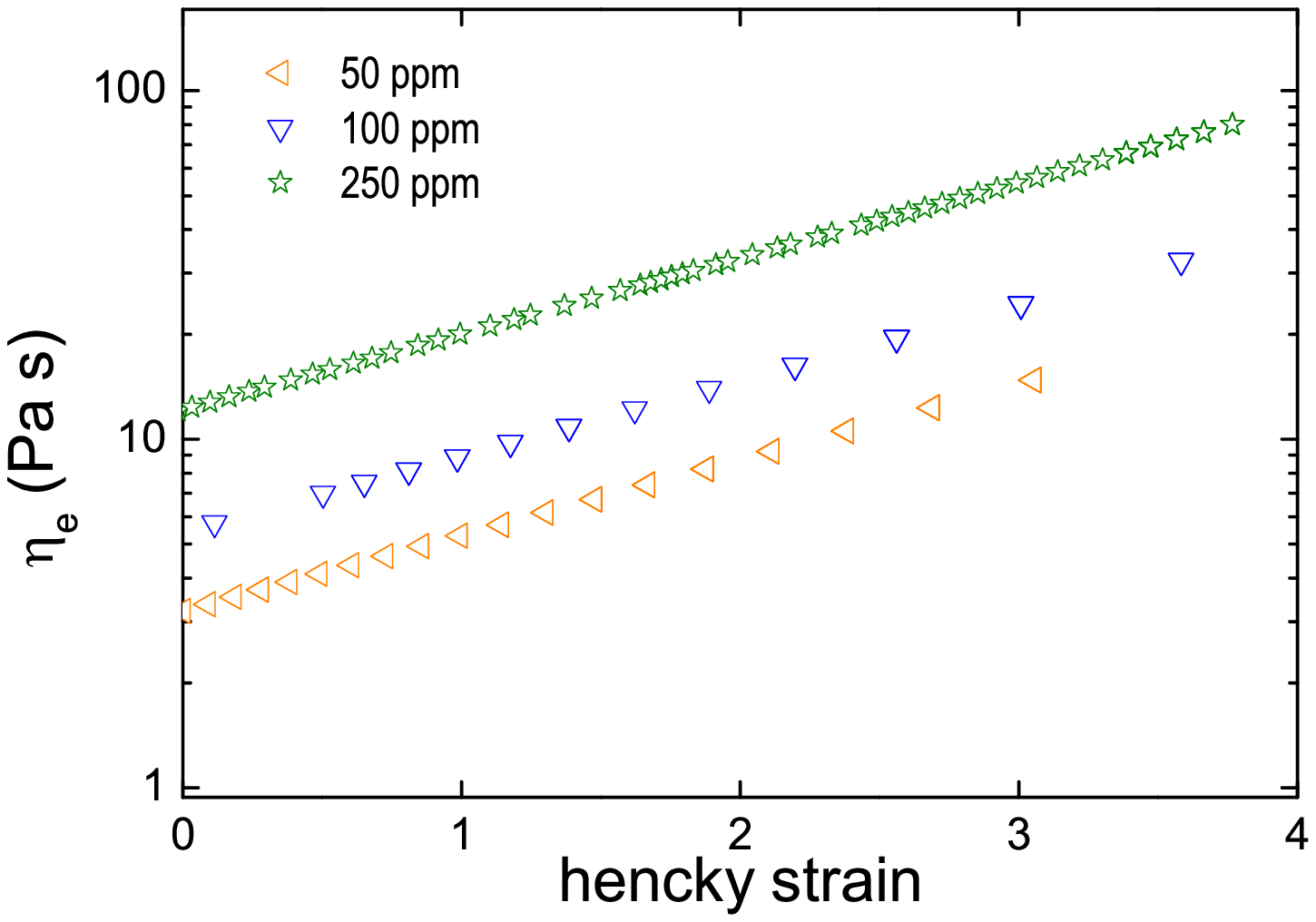}
\caption{Left) The minimum neck diameter $h_{min}$ vs. time for
water ($\circ$) and 100ppm PEO($M_w=4*10^6$~g/mol) in water
($\Box$). Dotted line: 2/3 power law line as a guide for the eye.
Dashed line: exponential fit corresponding to the Rayleigh-Plateau
instability of a liquid column of water. Full line: exponential fit
in the regime of strongly increasing elongational viscosity. Right)
The elongational viscosity for different concentrations of PEO as a
function of the Hencky strain. } \label{eloviscohenky}
\end{figure}

In the experiments presented here (fig\ref{figdroplets}), solutions of the flexible polymer PEO at
low concentrations (10-2000 ppm) in a low viscosity solvent were used. In these solutions, elastic
contributions at small deformations are weak and, as a consequence, the complex viscosity $\eta^*$ is difficult to determine by small amplitude oscillatory shear rheometry. This means that, for small deformation, the polymers do not affect the flow. Indeed, the experimental results (fig\ref{eloviscohenky}) show that the dynamics of the primary stages of
the detachment process are not altered. Only when the flow (and thus
the elongation) is strong enough do the polymers become stretched, interrupting the
finite time singularity of the pinch-off process. Instead, one observes the formation of a filament and an abrupt transition to a new exponential regime with a much larger time scale. This inhibition of the finite time singularity was first observed by Amarouchene et al. \cite{Amarouchene2001}.

This is a surprising observation because it means that the elastic
stresses in the liquid that balance the stresses from the surface
tension are much higher than for the case of the pure solvent (fig.
\ref{eloviscohenky}). Still, the shear viscosity of the sample is
close to the solvent viscosity, even at high shear rates. The
solution to this apparent contradiction is the following: any type
of flow can be divided into a rotational and an elongational part.
The elongational part stretches the polymers and induces stresses.
The rotational part, e.g., in shear flow, causes the polymers to tumble
and stresses are averaged out to a large degree. In the filament,
the flow is purely elongational and stretches the polymers most
efficiently. This leads to the so-called elongational viscosity
$\eta_e$. For Newtonian liquids, the elongational viscosity $\eta_e$
is directly given by geometrical considerations and it follows
$\eta_e = 3 \eta_{shear}$. The factor $3$ is called the Trouton
ratio. For polymers, the situation is less clear and measurements of
the elongational viscosity are non-trivial. The experiment now gives
us the opportunity to estimate $\eta_e$. In fact, a similar method
is used in a commercialized version of the experiment, the so-called
CaBER (Capillary Break up Extensional Rheology, Thermo-Fisher
Scientific, Karlsruhe, Germany).

 Instead of a falling droplet, CaBER uses a capillary bridge formed
between two plates that are separated
abruptly until the capillary bridge starts to contract. The force
balance of the surface stresses and viscous stresses yields an
apparent elongational viscosity
$\eta_e=\frac{\sigma}{r(t)\dot\epsilon}$, with $\sigma$ the surface
tension \cite{Stelter2001}. Indeed, if the filament is cylindrical, the
elongational rate can be obtained via $\dot\epsilon =
\frac{\dot r(t)}{r(t)}$ and it is constant for the exponential decrease
of the radius found in this regime. The minimal neck radius $r(t)$
shrinks exponentially with time and the elongational viscosity
$\eta_{e}$ thus increases exponentially with time. This is a type of
long-lasting start-up situation in which the increase in stresses holds
until the polymers eventually become fully stretched. Fig.
\ref{eloviscohenky} shows the elongational viscosity as a function
of the Hencky strain $H$. The Hencky strain, $H$, is a measure of the
elongation and is given by $H=\int \dot\epsilon dt$. The
elongational viscosity of the polymer solutions varies from 1 to 100
Pas and is thus five orders of magnitude higher than the shear
viscosity (see figure \ref{rheoPEO}).

\begin{figure}
\includegraphics[width=95mm]{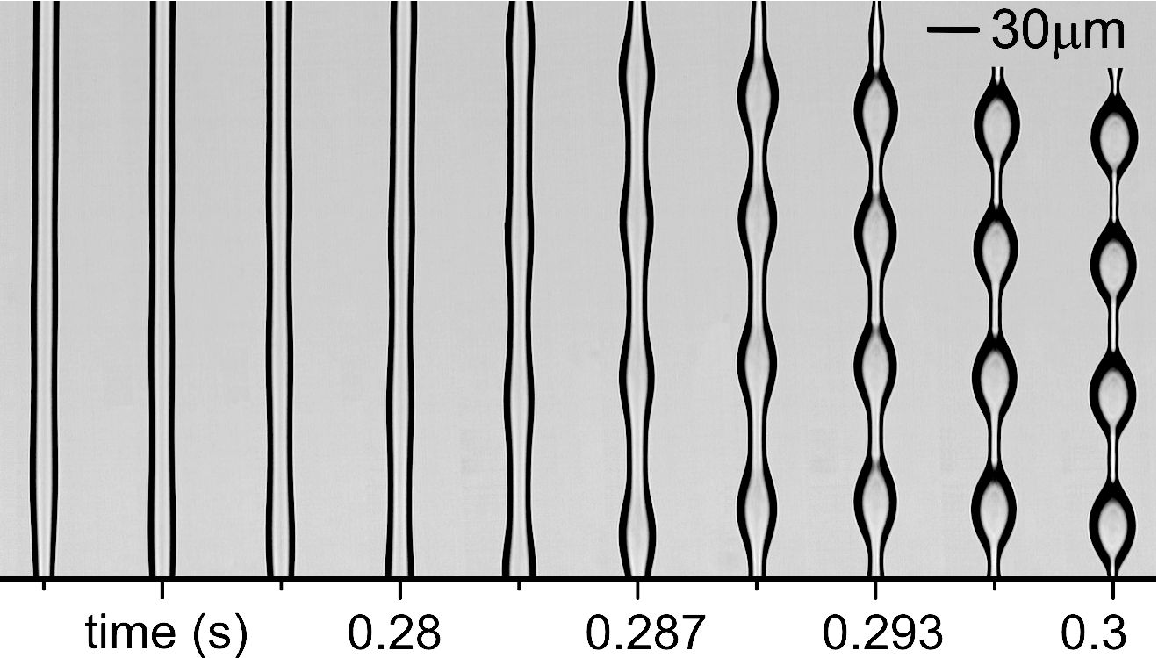}
\includegraphics[width=75mm]{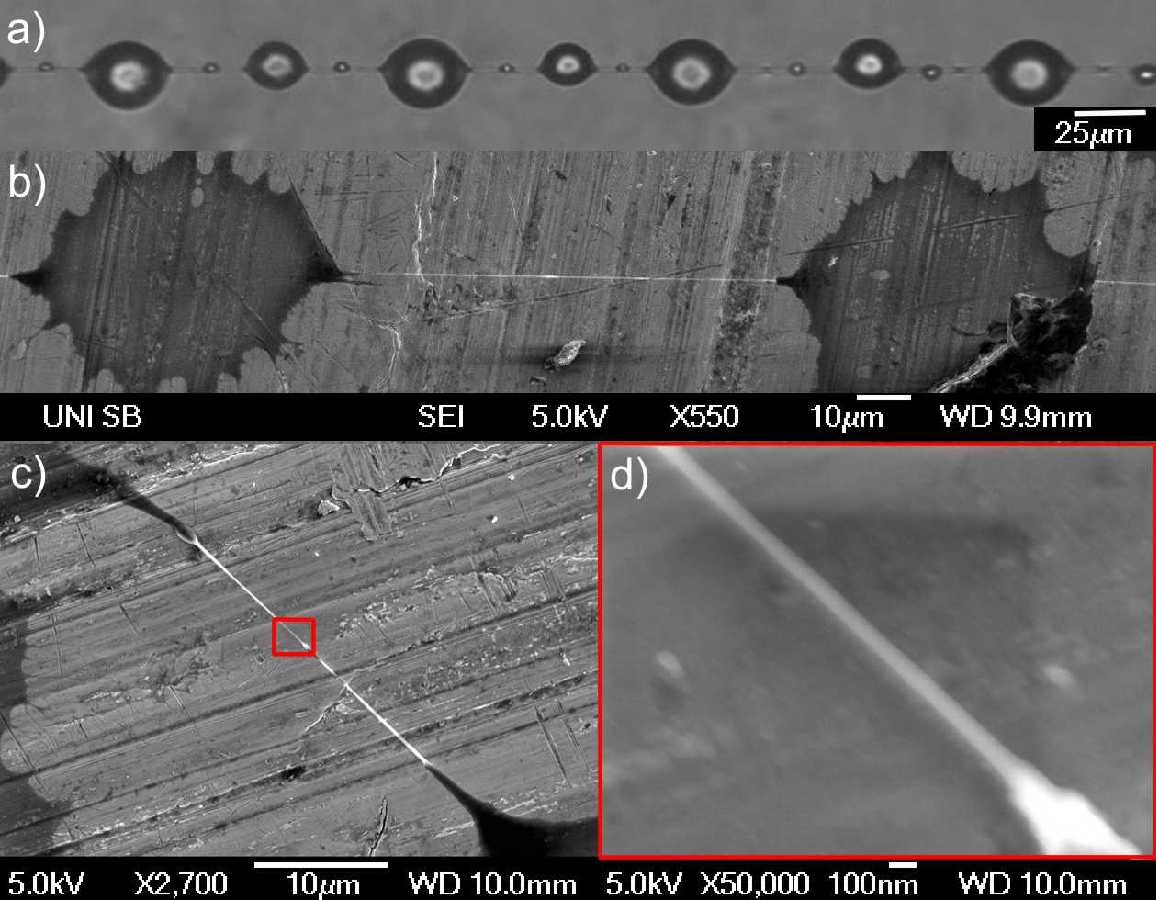}
\caption{Left) (Fig. 2 from \cite{Sattler2008}) Growth of a
sinusoidal instability of the viscoelastic filament of an aqueous
PEO solution (1000 ppm) that develops into a group of droplets on the
thinning filament. The spacing of the pictures is 1/300 s. Right)
(Fig. 4 from \cite{Sattler2008}) a) The final state of the filament.
Beads are formed off-center relative to the thread. b) Scanning
electron microscopy image of two beads connected by a thread
(intermediate resolution). The structure was caught and dried upon
the substrate. c) Another example of the structure; the red box
indicates a closeup at high magnification shown in d). The diameter
of the fiber can be as small as $70nm$.} \label{figrayleighplateau}
\end{figure}

The growing elastic stresses in the filament stabilize the flow, since
any distortions would lead to further stresses. This makes the
filament very robust and we might expect a Rayleigh-Plateau-like instability
to be observed only at the very end of the thinning process when the
polymers are fully stretched and elastic stresses cannot increase
further. However, in most of the experiments, a more or less
irregular instability scenario is observed at the very end of the thinning process and
singular "beads" grow on the filament. In a numerical study, Chang et al \cite{Chang1999}
predicted that the filament should start to disintegrate from both ends where it is connected to the falling droplet and the reservoir in the nozzle. At these points, the
curvature  and surface stresses are maximal. These authors predict an
iterative disintegration process, similar to the one
observed in more viscous Newtonian liquids. Experimentally, Oliveira
et al. \cite{Oliveira2005,Oliveira2006} found an iterative process
between generations of larger and smaller beads in a CaBER setup,
but their range of observations was limited.
Experimental data obtained by Sattler et al. \cite{Sattler2008} indeed revealed an instability process that was triggered from the ends; instead of an iterative process, we found
a Rayleigh--Plateau like instability scenario on the viscoelastic
filament. In contrast to previous studies, in which the plates where
pulled apart abruptly, in these experiments the separation was performed very gently in
order to prevent any additional distortions. Though in most of the
experiments singular droplets grew on the filament, it was possible
to observe the exponential growth of a coherent sinusoidal pattern,
(fig \ref{figrayleighplateau} right). The inverse growth rate of
the pattern was found to be $1/\omega = 9.3\pm0.1ms$. Linear
stability of a viscous fluid thread \cite{Eggers1997} predicts
$\omega=\gamma/(6R_{0}\eta_{eff})$, resulting in an estimated
extensional viscosity of $\eta_{eff} =9Pas \pm 2$, more than one
order of magnitude smaller than the extensional viscosity
$\eta_{E}\left({12\mu m}\right)=100Pas$ estimated above, but
nevertheless four orders of magnitude higher than the shear
viscosity. The exponential regime was followed by the final shrinking
of the remaining thread between the beads, which followed a linear law as
expected for highly viscous liquids \cite{Eggers1997,Rothert2003}.

We must still answer the question of how the filament finally
breaks. For polymer concentrations greater than 1000 ppm,
it was observed that the liquid filament connecting two beads did not
break; instead,  a pattern that resembles a solid fiber, as shown in
Fig.\ref{figrayleighplateau} right, a), was formed. The beads sit
alongside the filament; experiments on fluid drops on a fiber
\cite{Carroll1986} show that there must be a {\it finite} contact
angle between the drops and the filament for such a symmetry
breaking to occur. It follows that the thin filament must have formed a
(solid) phase different from that of the drops \cite{James1980}.
This hypothesis could be confirmed by the
scanning electron microscopy (SEM) images shown in
Fig.~\ref{figrayleighplateau} right b), in which remnants of two
droplets connected by a persistent thin thread are seen. Increased
magnification (panels c and d) allowed us to estimate the diameter
of the fiber as $75-150nm$. Our interpretation of this data is that, due to the coupling of stress
fluctuations and concentration fluctuations,  a flow-driven
phase separation takes place. Microscopically, one can imagine that
while solvent drains from the filament, the polymers become entangled, leading to even higher polymer concentration and increased entanglement, i.e., flow-induced phase separation takes place
\cite{Kume1997}. Further evidence for this concentration process was
found in \cite{Sattler2007}, where birefringence
measurements were performed to examine molecular conformations in
the break-up process. Evaporation can be excluded as a factor in
the formation of solid fibers, based on experiments in a two-fluid
system.

\section{Saffmann -- Taylor}
\label{ST}

\subsection{The classical Saffman -- Taylor instability}

The classical Saffman-Taylor instability occurs when, for example, air
pushes a viscous fluid in a narrow channel of height $b$ and
width $W$, a so called Hele--Shaw cell. In the following, the
viscosity of air is neglected and the viscosity of the viscous
liquid is given by $\eta$. The surface tension between the two
fluids is $\sigma$ and the viscous liquid is considered to perfectly
wet the channel.

\begin{figure}
\includegraphics[width=170mm]{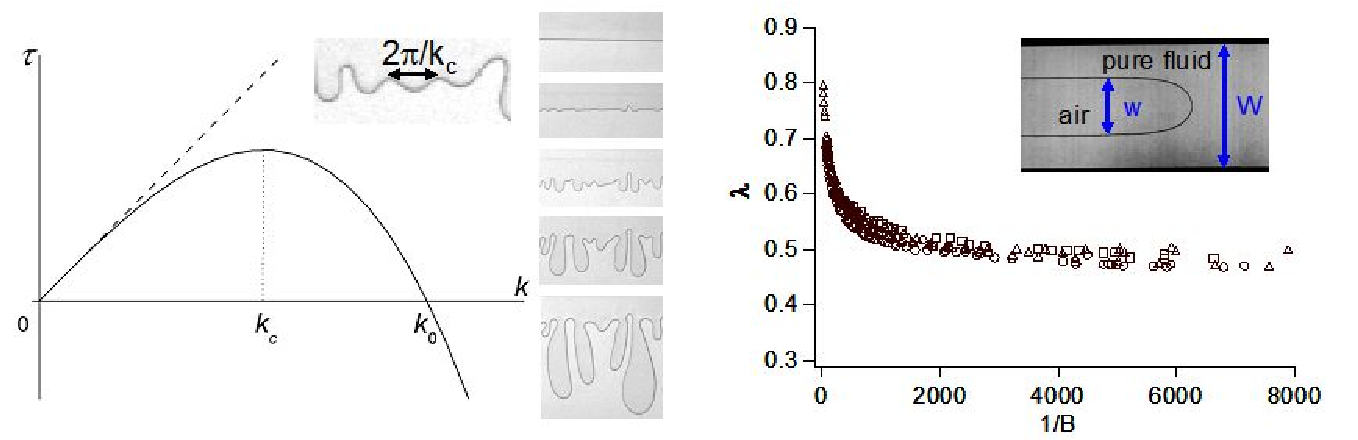}
\caption{Left) Growth rate $\tau$ as a function of the wave number
$k$. The most unstable wavelength $l_c$ is given by $2\pi/k_c$.
Inset: Snapshots of the destabilization of a planar front between
air and silicon oil (courtesy D. Derks and A. Lindner). Right)
relative finger width $\lambda$ as a function of the control
parameter $1/B$ for air pushing silicon oils of different viscosities
in channels of different geometries. Inset: snapshot of a finger
advancing into a linear cell (courtesy C. Chevalier).}
 \label{fig:STclass}
\end{figure}

Flow in the confined geometry is then governed by Darcy's law, which
gives the mean velocity (averaged over the thickness of the channel)
of the fluid as a function of an applied pressure gradient : ${\bf
V}= -\frac{b}{12 \eta}\mbox{\boldmath $\nabla$\unboldmath} p$. The
incompressibility of the fluid reads $\nabla \cdot {\bf V}=0$ and one
thus deals with growth in a Laplacian pressure field
$\mbox{\boldmath $\Delta$} p=0$. The pressure jump at the interface
is given by $\delta p=\sigma \left(\frac{2}{b}+\kappa \right)$, with
$\kappa$ being the curvature in the direction of the channel width,
once again using a two-dimensional approximation. Together with the
boundary conditions, this set of equations
completely determines the problem.

When the less viscous fluid pushes the more viscous fluid, an
initially straight interface becomes unstable. Small perturbations lead
to an increased pressure gradient and a higher velocity in
front of the perturbations and are thus amplified. Surface tension,
on the other hand, stabilizes the initially straight interface. The
competition between viscous and capillary forces leads to the
emergence of a characteristic length scale that can be calculated
using linear stability analysis \cite{Chuoke1959}. The maximum growth
rate is found for a wavelength $l_c=\pi b/\sqrt{Ca}$ with
capillary number $Ca=\eta U/\sigma$. The small fingers grow and
begin to compete with the more advanced fingers, screening the less
advanced fingers. In a linear channel of width $W$, one finally
observes a single finger propagating through the cell, the result of
a non-linear growth process. An example of initial finger
growth and finger competition, together with the growth rate obtained from the linear
stability analysis, can be seen in figure \ref{fig:STclass}a. The relative width of the single finger
$\lambda$, defined as the ratio between the finger width $w$ and the
cell width $W$, is given by the control parameter $1/B=12 Ca
(W/b)^{2}$; that is, the ratio between the two length scales of the
system $W$ and $l_c$.  Representation of the results obtained using
different fluids (and thus different surface tensions or different
viscosities) and different cell geometries as a function of the
control parameter $1/B$ shows that the results fall on a universal curve (see
figure \ref{fig:STclass}b). With increasing velocity $U$ of the
finger tip, viscous forces become increasingly important compared to
capillary forces and the relative finger width decreases. At high
velocity, the finger width does not, however, tend to zero but
stabilizes near a plateau value at $\lambda=0.5$.

This instability was described by Saffman and Taylor
\cite{Saffman1958} in 1958; however, finger selection remained a
puzzle for several decades. Neglecting surface tension, Saffman and
Taylor found a family of analytical solutions of the shape of the
interface given by
$x=\frac{W(1-\lambda)}{2\pi}ln\left[\frac{1}{2}\left(1+\cos\frac{2\pi
y}{\lambda W}\right)\right]$ that agrees well with experimental
observations. This treatment does not, however, explain the selection of a given
finger width. For this, one must take the surface tension into
account. This was done numerically by McLean and Saffman in 1981
\cite{McLean1981}. The selection process was
solved analytically only much later \cite{Hong1986, Shraiman1986, Combescot1986} and
was attributed to the fact that the surface tension represents
a singular perturbation leading to a solvability condition at the
finger tip. It is this condition that selects the finger from the
family of solutions found by Saffman and Taylor.

So far, we have treated an ideal two-dimensional problem. In reality, however, there
is a thin wetting film that remains between the advancing
finger and the glass plates. The thickness of this film, according to the Bretherton law, is
proportional to $Ca^{2/3}$
$t/R=0.643(3Ca)^{2/3}$ \cite{Tabeling1986}. As a consequence, the
pressure jump at the interface is continuously modified. This three-dimensional
effect leads to the slight modification of the finger width observed
between different experimental geometries, as observed in figure
\ref{fig:STclass}.

\subsection{Finger narrowing in shear thinning fluids}

\begin{figure}
\includegraphics[width=90mm]{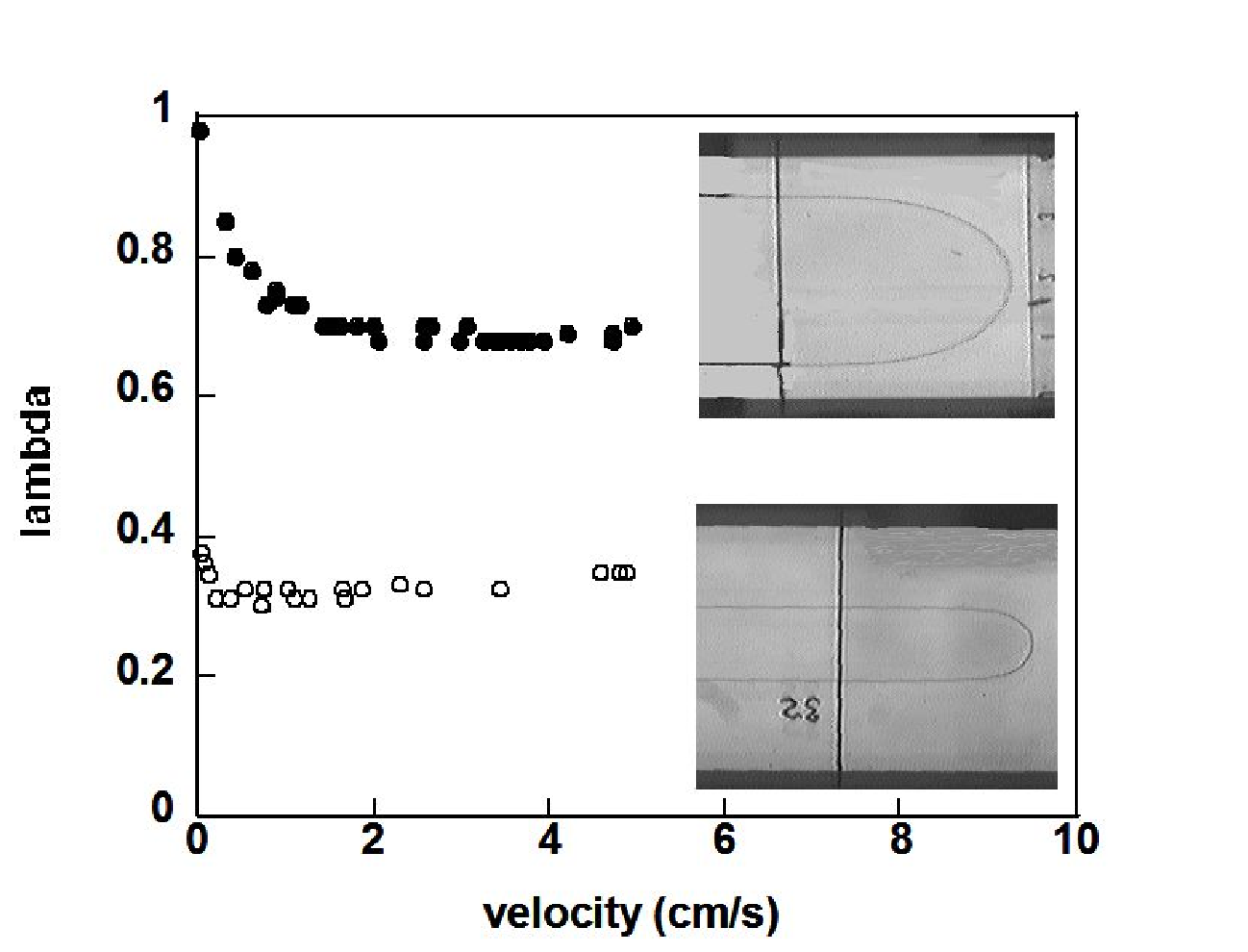} \hspace{0.5cm}
\includegraphics[width=60mm]{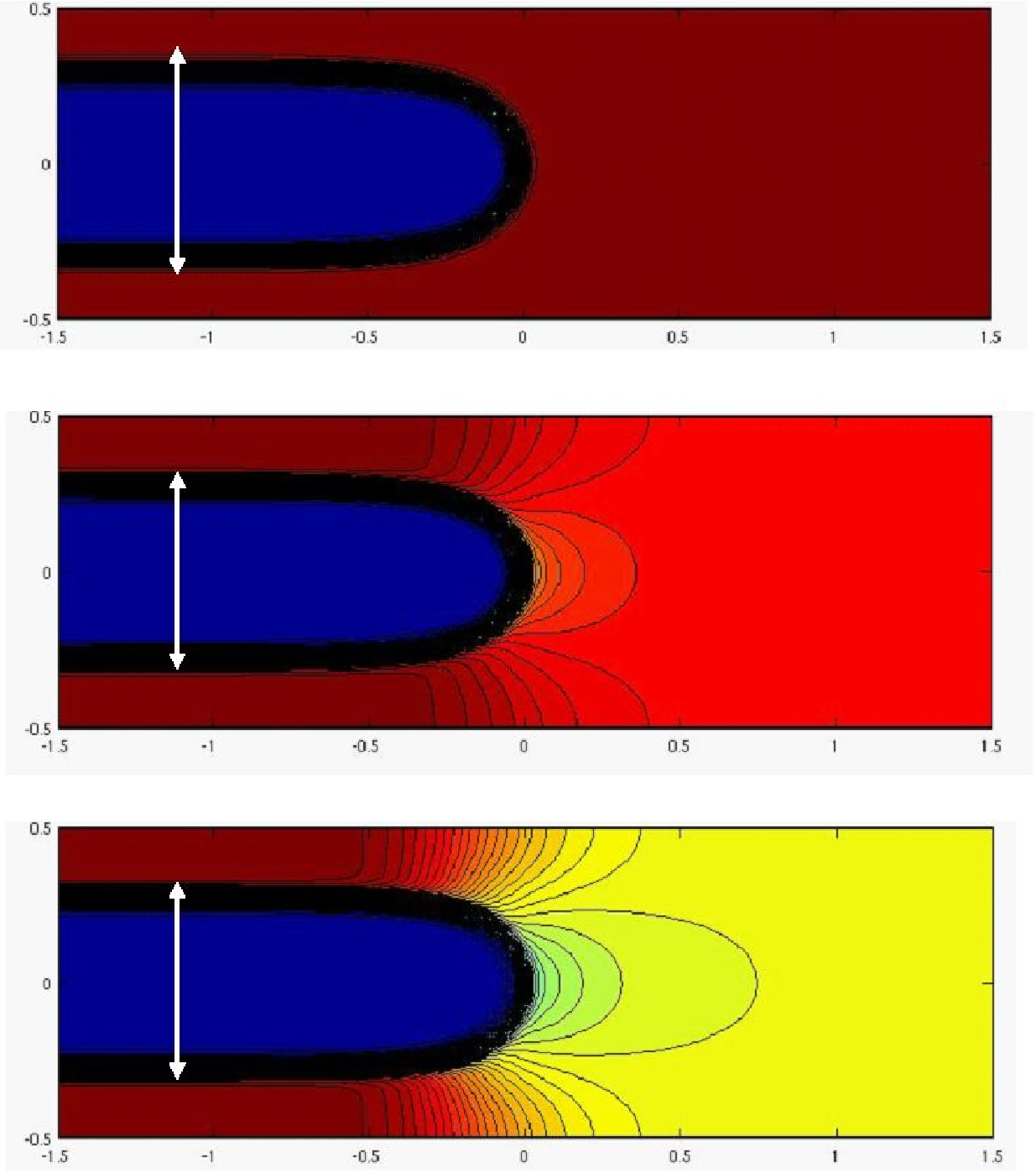}
\caption{Left) Relative finger width $\lambda$ as a function of the
velocity $v$ for solutions of xanthane ($\circ$) at 1000 ppm and PEO
($\bullet$) at 50 ppm. Inset: snapshots of fingers in the two
solutions at high velocity (4 cm/s). Right) Viscosity in front of
the advancing finger in a shear thinning fluid found from numerical
simulations \cite{Nguyen2009}. The shear thinning effect increases
from top to bottom. Red corresponds to high viscosities and yellow
to low viscosities.}
 \label{fig:Pictures}
\end{figure}

When performing experiments in dilute solutions of xanthane, one
observes a strong modification of the selection process. As can be
seen in the snapshot in Figure \ref{fig:Pictures}a, at high velocity, fingers are found
to be significantly narrower than the classical limit of
$\lambda=0.5$. This observation can be qualitatively
linked to the behavior of the shear thinning fluid pushed by
the finger in the Hele-Shaw cell. Numerical simulations
\cite{Nguyen2009,Kondic1998} show that the viscosity is not uniform
throughout the cell (see figure \ref{fig:Pictures}b); regions of
high fluid velocity and thus high shear rate have a low viscosity.
This is essentially the case in front of the finger tip and the
system becomes anisotropic, leading to finger narrowing.

For weak shear thinning ($1>n>0.65$), we have shown
\cite{Lindner2000a, Lindner2002} that simply replacing in the
control parameter $1/B$ the constant viscosity $\eta$ by a shear-
dependent viscosity $\eta(\g)$ allows rescaling the data onto the
universal curve for Newtonian fluids. The shear rate $\g$ is here
the average shear rate in the cell. For stronger shear-thinning
($n<0.65$), this rescaling fails and deviations from the classical
result toward smaller fingers are observed.

Narrower fingers have also been observed by Rabaud {\it et al.}
\cite{Rabaud1988}. Using a Hele -- Shaw cell with engraved glass
plates, they found viscous fingers with $\lambda$ significantly
smaller than $0.5$ for Newtonian fluids. The observation of such
"anomalous" fingers is explained by the fact that the engravings
represent a local perturbation at the finger tip. This disturbance
removes the classical selection of the discrete set of solutions.
The continuum of solutions given by the analytical result of Saffman
and Taylor without surface tension then becomes accessible:
$\lambda$ can take values smaller than $0.5$ at high velocity.
Rabaud {\it et al.} showed that for a given value of the capillary
number $Ca$, it is not the relative finger width that is selected but
that the dimensionless radius of curvature at the tip $\rho/b$. $\rho$
can be linked to the finger width $\lambda$ via the relation
$\rho=\frac{\lambda^2 W}{\pi (1-\lambda)}$, which follows from the finger
shape predicted by Saffman and Taylor.

\begin{figure}
\includegraphics[width=60mm]{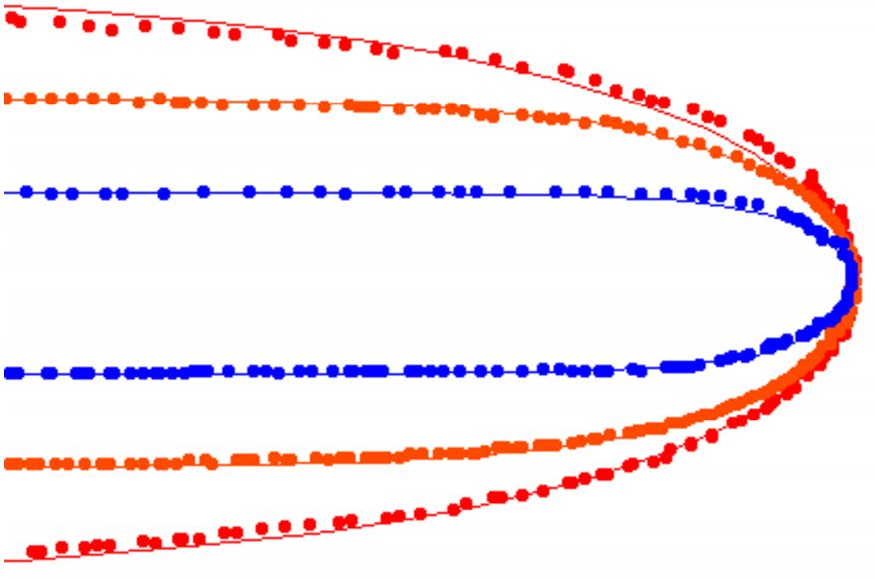}
\hspace{1.5cm}
\includegraphics[width=80mm]{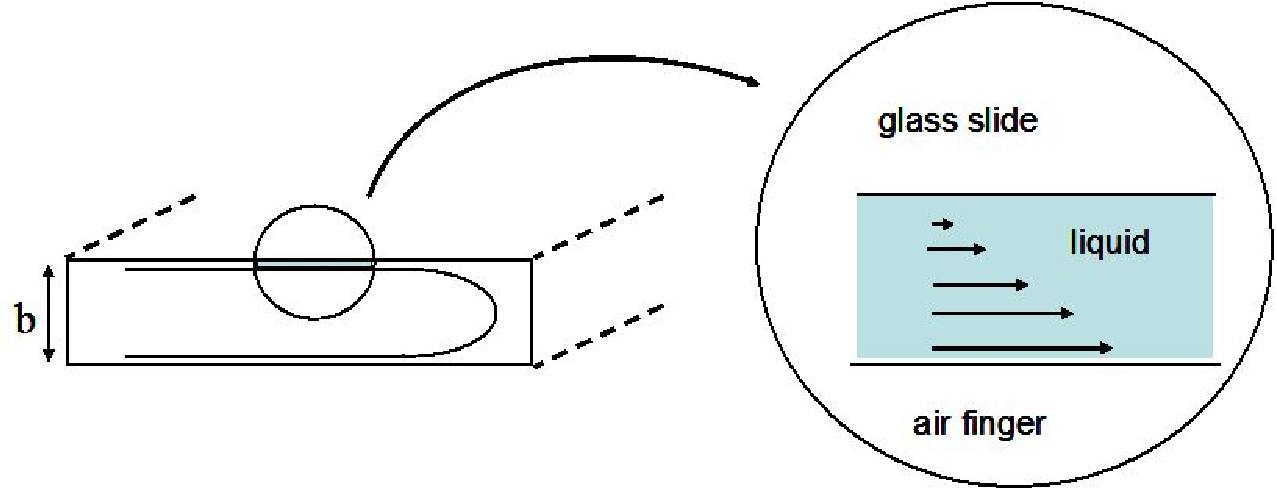}
\caption{Left) Experimental finger shapes in a solution of xanthane
of 2000~ ppm for three different channel widths $W$=2, 4, and 8~ cm.
Right) Sketch of the thin wetting layer observed between the finger
and the glass plates}
 \label{fig:shape}
\end{figure}

A similar mechanism has been found to be responsible for the
selection of the viscous fingers in a shear thinning fluid. Here
when the shear thinning character of the fluid is strong enough,
anisotropy plays the role of the perturbation at the finger tip.
Figure \ref{fig:shape}a shows experimental finger profiles for a
given capillary number and three different cell widths.  One clearly
observes that the radius at the finger tip is identical for the
three experiments, leading to lower finger width compared to the
Newtonian case, in which the relative finger width $\lambda$ is
selected. The relation between $\rho/b$ and $\lambda$ is found to
depend on the shear thinning character of the fluids and thus the
shear thinning exponent $n$. Knowledge of the relationship between $\rho/b$
and $\lambda$ solves the selection problem, as we can now predict the
finger width $\lambda$ from the rheological data. The presence of
shear thinning thus leads to a completely different selection
mechanism that is closer to what is observed, for example, in dendritic
growth \cite{Couder2000} and which requires anisotropy in the system. Corvera
Poir\'e {\it et. al.} \cite{Corvera1998} directly solved the problem
for a power law fluid; their results are in good agreement with
the experimental observations. Note that at high
velocities one observes a saturation of $\rho$, leading to an increase of the
finger widths. This might be attributed to inertial effects, which
have been observed to increase finger width \cite{Chevalier2006}
and which may begin to play a role in this low viscosity fluid.

\subsection{Finger widening due to normal stresses}

In experiments using solutions of the flexible polymer PEO,
completely different behavior is found. In contrast to the
observations in shear thinning fluids, where finger narrowing
occurs, one now observes finger widening compared to the Newtonian
case (see figure \ref{fig:Pictures}a).

The presence of normal stresses in the thin wetting layer might be
responsible for the finger widening; one can attempt to account for
this effect by adding a supplementary pressure to the system.  In
classical theory, the pressure jump at the interface between two
liquids is given by the radius of curvature. Tabeling {\it et
al.} \cite{Tabeling1986} have shown that one can incorporate the
effect of a finite thickness of the wetting film by correcting the
surface tension in the control parameter. Following the same
argument, in the control parameter one can add the supplementary
pressure caused by the normal stresses to the surface tension term
$\sigma*=\sigma+1/2 N_1(\g)b$. For moderate normal stresses, this
 allows rescaling of the data onto one universal curve and once
again solves the selection problem.

\section{Faraday instability}
\label{Faraday}

\subsection{History}
\label{Faraday-history}

The Faraday experiment was first reported in 1831
\cite{Faraday1831}. In an appendix of a paper on Chladni
figures, Faraday  reports on the crispated state of a layer
of liquid that is shaken vertically. He refers to works by
\emph{"Oersted, Wheatstone and Weber and probably others (sic)" } who
had earlier mentioned the phenomenon; however, according to Faraday, it was
he who gave the first conclusive description. The experimental
setup consisted of a box that was mounted on a rod. The rod was set
into vibration by a bow; its oscillation frequency was presumably
on the order of a few Hz. When the oscillation amplitude exceeded a
certain critical value, standing capillary surface waves were
observed. Apparently, Faraday was impressed by the richness of the
patterns and noted: \emph{"obtained in this way the appearances were
very beautiful, and the facilities very great"}. It is a remarkable
achievement that even at that early date Faraday found that the surface waves
oscillate with half of the driving frequency. This is the so-called
subharmonic response. Faraday also pointed out
differences in the wavy surface patterns when he compared simple oils
with, e.g., the white of an egg. He stated: \emph{"The difference
between oil and white of egg is remarkable; . . . the crispated
state may be a useful and even important indication of the internal
constitution of different fluids."}

\begin{figure}
\begin{center}
\includegraphics[width=150mm]{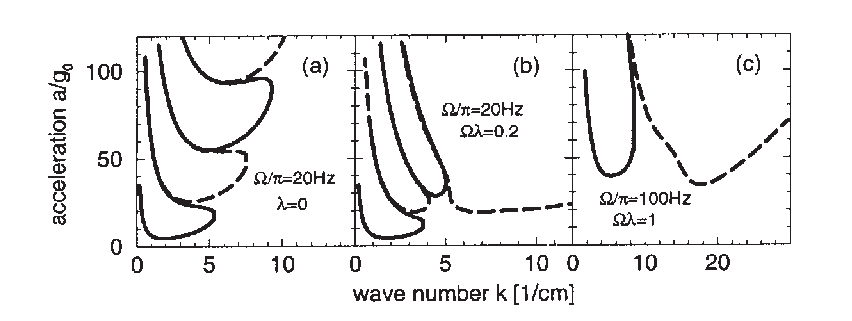}
\caption{(From \cite{Muller1999}) The linear stability diagram of
the Faraday experiment for a Maxwell fluid. Full lines mark the
transition to the subharmonic response and dashed lines to the
harmonic one. a) For a Newtonian fluid, i.e., when the polymer
relaxation time $\tau_p$ is zero, the subharmonic response becomes
unstable first. b) For finite relaxation times, a new harmonic tongue
appears c) Only when the inverse polymer relaxation time compares to
the driving angular frequency $\Omega$ does the new harmonic instability
tongue become unstable first.} \label{figfaradayzungen}
\end{center}
\end{figure}

In 1868, Matthiesen \cite{Matthiesen1870} reported on systematic
measurements; he stated incorrectly that the surface response
should be synchronous to the drive. In 1883, Lord Rayleigh
\cite{Rayleigh1883} proposed a theoretical treatment in terms of a
parametric pendulum, the Mathieu oscillator. In 1954, Benjamin and
Ursell \cite{Benjamin1954} solved the linear problem for ideal
liquids (without viscosity) with an infinite set of Mathieu
oscillators that oscillate with integral and $(n+1)/2$ multiples of
the driving frequency. The integral multiples correspond to a
possible harmonic (synchronous) response and the $(n+1)/2$ to a
subharmonic response (figure \ref{figfaradayzungen}). In 1994, Kumar and
Tuckermann \cite{Kumar1994} presented a numerical analysis of the
linear problem in the case of viscous liquids with finite depth of the layer.
Together with an analytical treatment, this analysis was used to find
parameters to experimentally reproduce a harmonic response
using a very thin layer of liquid
\cite{Muller1997}.
   Recently, a large variety of patterns with
up to 10-fold rotational symmetry (quasiperiodic), superlattices or
localized patterns have been reported
\cite{Kudrolli1998,Binks1997,Arbell1998}. A variety of these
patterns could be obtained using a simple liquid driven
with a single frequency. Near onset, in linear order, a single wave
number first became unstable. The resulting wave vector(s) could
have any orientation, but nonlinear interaction with the higher
harmonics led to a given pattern selection process
\cite{Zhang1996,Muller1997,Edwards1994}. Obviously, the pattern
dynamic might become even richer if \emph{two} discrete wave numbers
become unstable simultaneously.

It is worth mentioning that numerical simulation
of the full hydrodynamical problem of the Faraday experiment with a
single driving frequency for simple liquids became available only very recently
\cite{Perinet2008}. P\'{e}rinet {\it et al.} solved the complete
nonlinear Navier -- Stokes equations by a finite-difference
projection method coupled to a Front Tracking technique for the
calculation of the surface tension forces and advection of the
interface. They compared the complete spatial and temporal Fourier
spectrum of the surface state and found good agreement with
experimental data from Kityk et al. \cite{Kityk2005}. A quantitative
theoretical description of the nonlinear wave state of complex
liquids remains an even more challenging problem that is still
unsolved.

\subsection{Complex liquids}

 \label{Faraday-complexliquids}

 The first experimental data
on Faraday waves of polymer solutions were, to our knowledge,
presented in 1998 by Raynal, Kumar and Fauve \cite{Raynal1999}.
Their work concentrates on \emph{dilute} polymer solutions where the
influence of elasticity is small. They found a slight shift of the
critical acceleration; the critical wave numbers were not
affected. In 1999, M\"{u}ller and Zimmermann presented a linear
stability analysis for a Maxwell fluid \cite{Muller1999}. They found
that when the inverse of the relaxation time of the Maxwell fluid
compares to the driving frequency, a harmonic response might become
unstable first and that, for a certain set of parameters, a bistable
situation exists; see fig. \ref{figfaradayzungen}. Again, in 1999,
Wagner and M\"{u}ller presented experimental and theoretical data on a
Faraday experiment using a concentrated polymer solution of 2000 ppm
PAA ($M_w=4*10^6$~g/mol) in a water-glycerol mixture  \cite{Wagner1999} with a
relaxation time $\tau_p\sim 1$~s (see section \ref{HD}). The
rheological data obtained for the complex viscosity was then fed
into the linear stability algorithm of Kumar and Tuckerman
\cite{Kumar1994} and a reasonable agreement of the experimentally
and numerically determined critical accelerations was obtained.
Notably, a bicritical situation was found in which the subharmonic
and harmonic responses became unstable simultaneously.

\begin{figure}
\includegraphics[width=100 mm]{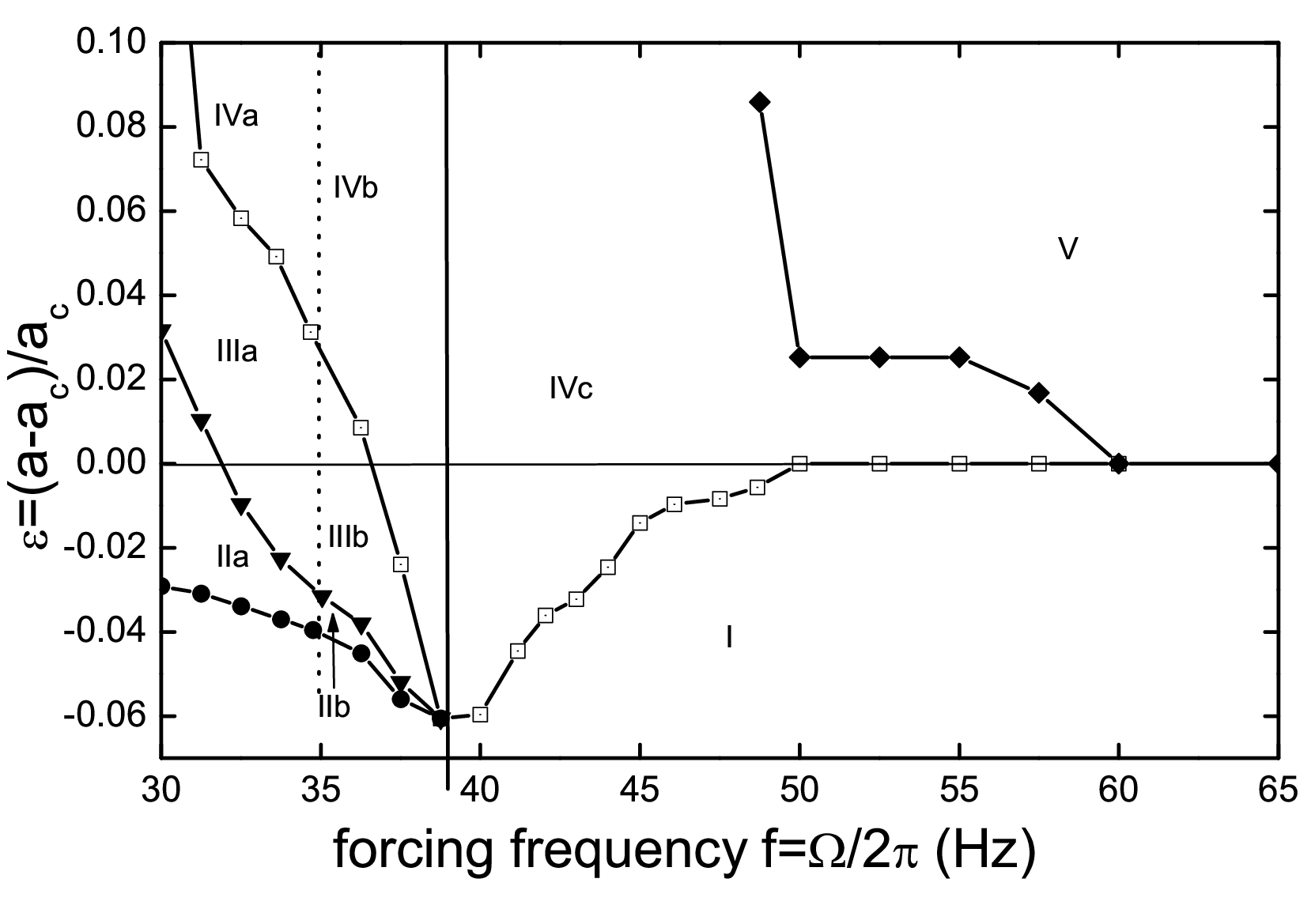}
\caption{(Fig. 2 from \cite{Wagner1999}) Phase diagram of the
observed nonlinear patterns. The symbols mark experimental data
points. The abscissa, $\epsilon=\frac{a-a_c}{a_c}=0$, indicates the
linear threshold with $a$ the acceleration amplitude and $a_c$ the
critical acceleration. The lowest solid line denotes the saddle
point of the hysteresis. Region I: flat surface; IIa: harmonic
hexagons covering the whole surface; IIIa: harmonic-subharmonic
hexagonal superlattice (see text) extending over the whole surface;
IIb, IIIb: as before, but patterns occur in the form of localized
patches surrounded by the flat surface; IVa, IVb, IVc: chaotic
dynamics of subharmonic lines competing with extended (a) or
localized (b) hexagonal superlattices or the flat surface (c); V:
stationary subharmonic lines extending over the whole surface
(Newtonian regime).} \label{phasediagram}
\end{figure}

\begin{figure}
\includegraphics[width=150mm,]{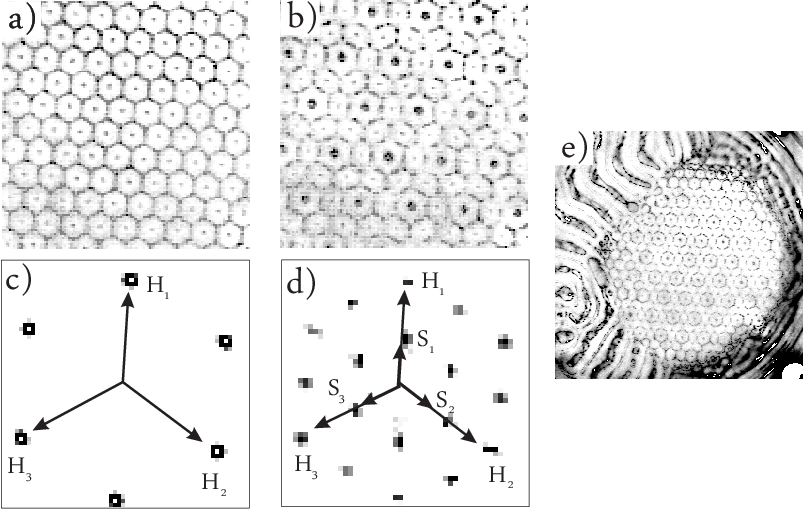}
\caption{ (a) The harmonic hexagonal pattern. (b) the
harmonic-subharmonic hexagonal superlattice. (c) and (d) the
respective spatial Fourier spectra. (e) (Fig. 4 from
\cite{Wagner1999}) Localized stationary surface patterns of harmonic
hexagons in coexistence with a localized nonstationary patch of
lines. } \label{figfaradaymuster}
\end{figure}

In fig. \ref{phasediagram}, an overview of the observed patterns is
presented. This phase diagram was obtained by setting a fixed frequency
and varying the driving strengths. When a (new) pattern was visible,
the transition point was noted. The primary instability was found to
be hysteretic ($\epsilon<0$) and most of the transitions between the
patterns were hysteretic as well (not shown). For frequencies well
above the bicritical frequency $f_b\approx 39 Hz$, a subharmonic
pattern of lines (region $V$) was observed corresponding to
observations in a Newtonian liquid of similar viscosity. For
parameters where only a harmonic response existed, a coherent pattern
of hexagons was observed (region $IIa$, fig.
\ref{figfaradaymuster}a). The hexagonal symmetry is generic for a
harmonic response where the temporal symmetry allows the coupling of
three wave vectors $H_n$ that are equally distributed on the
critical circle (fig.\ref{figfaradaymuster}c). For frequencies close
to the bicritical point, with increasing driving strength
subharmonic wave vectors $S_n$ also become unstable. They are slaved
by the harmonic hexagonal pattern and arrange together to a $2
\times 2$ superlattice. The nomenclature is taken from
crystallography and relates to the ratio of $1:2$ of the harmonic
and subharmonic wave vectors, which is close to the ratio of linear
unstable wave numbers (region $IIIa$, fig. \ref{figfaradaymuster}b).

The pattern-forming process can be understood in the following way.
The linear stability analysis reveals that the surface state consists
of a singular wave number $\emph{k}$ (only exactly at the bicritical
point do two wave modes synchronously become unstable) but of an infinite
series of temporal Fourier components, $n\Omega$ or $(n+1)/2\Omega$
for the harmonic or subharmonic responses, respectively. This linear
result is fed into the nonlinear equations for the hydrodynamic
velocity field $\textbf{v}$. The solvability condition in quadratic
order implies that resonant terms must vanish in the higher orders
to prevent secular growth. For the temporal components, an arbitrary
quadratic nonlinearity results in a frequency spectrum of integral
multiples of $\Omega$, whether or not S or H are considered. Thus,
quadratic nonlinearities are able to resonate  with {\em harmonic}
linear eigenmodes, but not with {\em subharmonic} ones. In the same
way, spatial resonance must be guarantied as well. Now,  any triplet of
harmonic modes $\{{\bf k}_{H1},{\bf k}_{H2},{\bf k}_{H3}\}$ with
$|{\bf k}_{Hm}|=k_H$ and ${\bf k}_{H1}+{\bf k}_{H2}+{\bf k}_{H3}=0$
are in resonance. This generic 3-wave vector coupling is well known
(e.g., from non-Boussinesq-Rayleigh-B\'{e}nard convection) and
enforces a saddle node bifurcation towards hexagonal patterns. The
associated solvability condition is referred to as the amplitude equation or
Ginzburg--Landau equation. Within the subharmonic regime, such a
resonant 3-wave vector coupling is prohibited due to the missing
temporal resonance. The pattern selection mechanism is
therefore then controlled by the cubic coupling coefficient in the associated
amplitude equations \cite{Edwards1994,Chen1997}, and a variety of
patterns is allowed.

 The secondary superstructure must result from a nonlinear excitation process,
as well. The responsible mechanism is again a 3-wave vector interaction. The linear
 analysis yields the result that the wave number of the subharmonic response is approximately
 $|{\bf k}_{S1}|=\frac{1}{2} |{\bf k}_{H1}|$. Invoking basic resonance arguments,
 one finds directly ${\bf k}_{S1}=\frac{1}{2} {\bf k}_{H1}$.

 For increased driving strength, localized patches of hexagons have also been observed.
 They cannot be explained by a triplet of real Landau equations supplemented by
 diffusive spatial derivatives, since within this familiar amplitude equation model,
 {\em stable} isolated islands of hexagons do not exist. For even larger driving
 strength, the patterns become chaotically time-dependent. Patches of
 subharmonically oscillating lines originating in an erratic manner
 from the cell boundary or the flat surface penetrate into the stationary hexagonal
 superlattice. They then disappear and the original structure is recovered.
 This process repeats itself on time scales of seconds to minutes, leading to a
 temporary coexistence  of the stationary hexagonal superlattice with subharmonic lines
 (region $IVb$, fig.~\ref{figfaradaymuster}). Higher driving amplitudes lead to a
 fully chaotic surface pattern. Still, no satisfying theoretical description of these states exists.

\section{Summary}

In the present paper, we have revisited three classical surface
instabilities, the Rayleigh--Plateau instability followed by the
detachment of a droplet, the Saffman--Taylor instability and the
Faraday experiment. We have described the modifications of the
selection processes that occur when polymer solutions instead of
simple Newtonian fluids are used. We have shown that the non-Newtonian
properties that come into play when using viscoelastic fluids
can fundamentally modify the selection processes. The
Rayleigh--Plateau instability, which describes the onset of the
detachment of a droplet, is not modified by the use of dilute
solutions of a flexible polymer, as the linear viscoelastic
properties of this solution are weak. The final detachment of the
droplet is, however, strongly delayed as high stresses develop when
the fluid filament becomes more and more stretched. The additional
resistance to breakup can be attributed to large elongational
viscosities present in these solutions. Only in the final stages of
the detachment process, when the polymers are fully stretched, might another
Rayleigh--Plateau instability be observed. Eventually, a flow-induced phase
separation may lead to a solid nanofiber. Using this
type of polymer solution also leads to a modification of the
Saffman--Taylor or viscous fingering instability. Normal stresses
develop in the thin wetting film left between the advancing finger
and the glass plates of a Hele--Shaw cell, leading to finger
widening compared to the classical results. When using shear
thinning solutions of a rigid polymer, finger narrowing is
observed. This can be attributed to an anisotropy that develops in
the system. In front of the finger tip, where the shear rate is
high, the viscosity is decreased. For strong shear thinning, the
selection process is quantitatively modified as the anisotropy
removes the solvability condition at the finger tip that is
responsible for the selection process in Newtonian fluids. The
selection process resulting from this anisotropy is found to be
closer to that involved in the formation of dendrites. The Faraday instability is
studied in a concentrated polymer solution that can be approximately
described by a Maxwell model. It has been shown that when the
relaxation time of the complex fluid is of the same order of
magnitude as the driving frequency, a subharmonic and an harmonic
response can become unstable simultaneously. This leads to an entirely
new class of patterns, the so-called superlattices.

\appendix

\section{Acknowledgements}
\label{}

We thank Julia Nase for a critical reading of the manuscript.

\bibliographystyle{elsarticle-num}

\end{document}